\documentclass[twocolumn,aps,pra,showpacs,superscriptaddress,floatfix]{revtex4}

\usepackage{amsmath}
\usepackage{amssymb}
\usepackage{amsfonts}
\usepackage{bm}
\usepackage{bbm}
\usepackage{epsf}
\usepackage{times}
\usepackage{nicefrac}
\usepackage{color}
\usepackage{graphicx}

\newcommand{\addrHD}{Max--Planck--Institut f\"ur Kernphysik,
Saupfercheckweg 1, 69117 Heidelberg, Germany}

\newcommand{\addrNIST}{National Institute of Standards and Technology,
Gaithersburg, Maryland 20899--8401}

\begin{document}

\sloppy

\title{Calculation of Hydrogenic Bethe Logarithms for Rydberg States}

\author{Ulrich D.~Jentschura}
\affiliation{\addrHD}
\affiliation{\addrNIST}

\author{Peter J.~Mohr}
\affiliation{\addrNIST}


\date{\today}

\begin{abstract}
We describe the calculation of hydrogenic (one-loop) 
Bethe logarithms for all states with principal quantum 
numbers $n\leq 200$. While, in principle, 
the calculation of the Bethe logarithm
is a rather easy computational problem involving only 
the nonrelativistic (Schr\"{o}dinger) theory of the 
hydrogen atom, certain calculational difficulties 
affect highly excited states,
and in particular states for which the principal
quantum number is much larger than the 
orbital angular momentum quantum number.
Two evaluation methods are contrasted. One of these is based on the 
calculation of the principal
value of a specific integral over a virtual photon 
energy. The other method relies directly on the spectral 
representation of the Schr\"{o}dinger--Coulomb propagator.
Selected numerical results are presented. The full set 
of values is available at arXiv.org/quant-ph/0504002.
\end{abstract}

\maketitle

\widetext


\section{Introduction}

The evaluation of the Bethe logarithm, in 1947~\cite{Be1947},
was carried out using one of the first automatized devices for 
the implementation of numerical calculations in physics.
Today, the evaluation of the (one-loop) Bethe logarithm for the 
ground state of hydrogen to about 10 figures of accuracy
can be carried in less than a second
on a modern workstation. 
Consequently, one might be tempted ask why 
there should be yet another paper on Bethe logarithms in the first place? 
The answer is threefold: (i) 
In the context of recent efforts toward an improved understanding
of the hydrogen and deuterium spectra~\cite{JeKoLBMoTa2005}
(see also physics.nist.gov/hdel),
we have striven to increase the number of states for
which this basic quantum electrodynamic correction
is known. (ii) The Bethe logarithm has been found to 
follow a characteristic asymptotic structure, expressible in terms
of a series in inverse powers of the principal quantum numbers.
This asymptotic structure 
has been found to be applicable to wide classes of quantum
electrodynamic effects in atoms~\cite{Je2003jpa,Je2004b60,JeEtAl2005}.
Consequently, it appeared to be of interest to verify these 
asymptotic properties by explicit calculations of the 
Bethe logarithm for very highly 
excited states. (iii) The most comprehensive collection of 
Bethe logarithms recorded so far in the 
literature~\cite{DrSw1990} extends up to 
the principal quantum number $n = 20$.
Here, we consider levels up to $n = 200$.
This should be contrasted with recent experimental 
investigations~\cite{dV2002} that were carried out with
states of principal quantum numbers as high as $n = 30$.

Frequency combs can lead to 
tremendous simplifications for high-precision spectroscopic 
experiments (e.g. Ref.~\cite{HuEtAl1998,HaEtAl1999,%
UdEtAl1999,ReEtAl2000,DiEtAl2000,JoEtAl2000,StTaHaWeTe2001,StScTaTe2002}).
In the future, it should become feasible 
to carry out high-precision experiments on transitions
much more effectively than in the past.
Rydberg states with  long natural lifetimes are rather promising
candidates for precision metrology, and one might imagine either 
direct transitions among Rydberg states or optical transitions
from, e.g., the metastable $2S$ state to a highly excited $D$ state.
Such measurements might contribute to future advances in our  
knowledge of the hydrogen and deuterium spectra.
On the theoretical side, the method of least squares~\cite{JeKoLBMoTa2005}
allows for a self-consistent adjustment of fundamental 
constants such as the Rydberg constant and the proton charge radius,
using experimental input data from more than one transition.
The investigation reported here is an element of a 
project (see also Ref.~\cite{JeKoLBMoTa2005})
to enlarge the 2002 adjustment of constants~\cite{MoTa2005}
so as to provide optimal predictions for energy levels
not contained in this adjustment, consistent with the 
values of the constants used in that adjustment.

Thus, we here
discuss the evaluation of Bethe logarithms using two different
methods: 
\begin{itemize}
\item an integral representation which is based on
analytic calculations using
the Sturmian representation of the Schr\"{o}dinger--Coulomb
Green function,
\item a spectral representation which relies on known 
results for the transition matrix elements of
discrete-discrete and discrete-continuum transitions  
of hydrogen.
\end{itemize}
The two methods are found to be suitable for different 
ranges of principal and angular momentum quantum numbers.
While a complete account of all previous work on the 
Bethe logarithm would result in an excessively long 
list of references, it might be 
instructive and appropriate to recall a few previous investigations 
on this subject~\cite{BeBrSt1950,Ha1956,ScTi1959,Li1968,%
Hu1969,KlMa1973,Sh1976,HaMo1985,Br1985,BaHiMo1989,FoHi1993}.
Recently, the Bethe logarithm has been re-evaluated,
for selected hydrogenic states, in the context
of lower-order terms acting as preparatory calculations
for higher-order relativistic corrections to the 
self-energy~\cite{Pa1993,JePa1996,JeSoMo1997}.
The basic equation defining the 
spectral decomposition has been given in Eq.~(2a),
and the basic equation for the integral representation 
has been indicated in Eq.~(2b) of Ref.~\cite{Br1985}. 
The evaluation for low-lying states of hydrogen, 
using the two methods, has previously
been discussed in Secs.~IIB (integral
representation) and  and Secs.~IIC (spectral representation) 
of the comprehensive Ref.~\cite{Br1985}.
A similar, though less comprehensive, comparison
of the two approaches had been made previously in Ref.~\cite{KlMa1973}.
In the context of the spectral representation,
we recall that in Table~II of Ref.~\cite{BeBrSt1950} and Table~I of Ref.~\cite{Ha1956}
one may even find results for the particular contributions of the 
discrete spectrum and of the continuum to the 
Bethe logarithm of selected low-lying states.
In general, we found that
unexpected numerical difficulties affect the calculation
of Bethe logarithms for Rydberg states,
especially in cases where the difference $n-l$
of the principal quantum number $n$ and the orbital
quantum number $l$ is large. 
Here, we attempt to enhance both the range of applicability of the 
integral representation as well as the spectral representation,
by the use of convergence acceleration 
methods~\cite{JeMoSoWe1999,AkSaJeBeSoMo2003} for the 
evaluation of hypergeometric functions that characterize propagator
matrix elements (integral representation) and the 
calculation of infinite sums over discrete virtual intermediate states
(spectral representation).

(At least) two further methods are 
available for the evaluation of Bethe logarithms:
one of these is based on the determination of 
approximate eigenfunctions obtained using finite basis
sets~\cite{HaMo1985}, which may be combined with a
Neville--Richardson extrapolation to yield accurate values,
thereby decreasing the required number of functions
in the basis set. Basis-set methods are also used
in calculations of Bethe logarithms in 
helium (see e.g. Ref.~\cite{Ko1999,DrGo2000}). A fourth method relies on a 
discrete-space (lattice) evaluation of the
radial component of the Schr\"{o}dinger--Coulomb 
propagator~\cite{JoBlSa1988,SaOe1989}.
This method is briefly discussed in Appendix~\ref{drep}.

A somewhat special role is played by
circular Rydberg states with $n-1 = l = |m|$~\cite{JeEtAl2005}, 
whose probability density around the atomic nucleus
approximately has the shape of a rotationally symmetric, ``circular'' 
tire (see Fig.~1 of Ref.~\cite{JeEtAl2005}),
but this shape is restricted to the 
highest possible magnetic angular momentum projection.
Because the Bethe logarithm does not 
depend on $m$, we will refer to all 
states with  $n-l = 1$ as circular states
in this article. Circular states have the highest
possible $l$ for given $n$. In the context of the 
current numerical investigation,
it thus appears useful to 
define a ``non-circularity'' or 
``angular-momentum defect'' $\zeta = n-l \geq 1$.
The radial hydrogenic wave functions have the structure 
of an exponential $\exp[-r/(n \, a_0)]$, 
where $a_0$ is the Bohr radius,
multiplied by a polynomial in $r$ with 
$\zeta$ terms.
The coefficients of this polynomial have an alternating sign 
pattern. Finite sums whose terms display an 
alternating sign pattern are notoriously problematic 
in numerical evaluations, because their convergence 
cannot be accelerated with the methods used for infinite
series, and the straightforward summation of the terms,
using multiprecision arithmetic, is often the only 
practical route to a reliable numerical evaluation.

This paper is organized as follows. In Sec.~\ref{trep},
we discuss the integral representation of the 
Bethe logarithm and its application to the 
calculation of levels with small $n-l$.
In Sec.~\ref{srep}, we discuss the spectral representation. 
Some brief conclusions are drawn in Sec.~\ref{conclu}.

%
%
\section{Integral Representation}
\label{trep}

As is customary for quantum electrodynamic 
bound-state calculations, we use 
a system of units in which $\hbar = c = \epsilon_0 = 1$.
The Bethe logarithm is a low-energy second-order 
perturbation which is 
due to virtual states with one photon mode excited,
and the atom in a virtual state. This can be seen 
most clearly by going to the Schr\"{o}dinger-picture representation
of the field operators~\cite{KuSt1995}.
A detailed discussion of the transition 
to the Schr\"{o}dinger picture for the
field operators, together with a basic application 
to bound-state problems, is given in Ref.~\cite{JeKe2004aop}.
The calculation naturally leads to 
an integral over the virtual photon energy
which involves a hydrogenic Green function with an argument
$z = E_n - \omega$, where $\omega$ is the energy of the virtual
photon. In the literature, it is customary to 
set $z = -(Z\alpha)^2 m/(2\,n^2\,t^2)$ (see, e.g., Ref.~\cite{Pa1993}).
So $t$ is a variable which parameterizes the argument of 
the Green function in terms of a generalized quantum number
``$n \to n\,t$.''
Of course, the variable $t$ has nothing to do with temporal
evolution. Solving for $t$, we obtain
\begin{equation}
t = \frac{1}{\sqrt{\displaystyle 
1 + \frac{2 \,n^2\,\omega}{(Z\alpha)^2 m}}}\,.
\end{equation}
Here, $m$ is the electron mass, $Z$ is the nuclear charge,
and $\alpha$ is the fine-structure constant.
The photon energy can be expressed in terms of $t$ as
\begin{equation}
\omega = \frac{1 - t^2}{t^2}\, \frac{(Z\alpha)^2 m}{2 n^2}\,.
\end{equation}
We denote the reference state by $|nlm\rangle$.
The relevant matrix element,
which involves the Schr\"{o}dinger--Coulomb propagator, is
given by
\begin{equation}
P_{nl}(t) = \frac{1}{3 m}\, \sum^3_{i=1}
\left< nlm \left| p^i \, \frac{1}{H - E + \omega(t)} \, p^i 
\right| nlm \right>\,.
\end{equation}
The integral representation for the Bethe logarithm
$\ln k_0(n,l)$, in terms of $P(t)$, is different
for $S$ states ($l=0$) in comparison to non-$S$ states
($l \neq 0$),
\begin{equation}
\label{tlnk0}
\ln k_0(n,l) = -  \frac{3}{4}\,\,
{\rm (P.V.)} \int_0^1 {\rm d}t\,\frac{1}{t^3} \, \left\{ 
\frac{t^2-1}{n \, t^2}\,P_{nl}(t) + \frac{2}{3\,n} - 
\frac{8}{3}\,t^2\,\delta_{l,0} \right\}  - 2 \,\ln(n) \,\,\delta_{l,0}\,.
\end{equation}
The specification of the principal value is necessary in 
Eq.~(\ref{tlnk0}) because of bound-state poles whose
residue gives the one-photon spontaneous decay width 
of an excited atomic state.

As an example, we consider here the $4P$ state for which the 
matrix element has the form
\begin{align}
\label{p4p}
P_{n=4,l=1}(t) =&
-\frac{1024 \, t^7}{45 \,(t-1)^8 \,(t+1)^8} \, \Phi(4,t) \,
\left[75 - 1700\,t^2 + 9954\,t^4 - 21124\,t^6 + 14907\,t^8\right]
\nonumber\\[2ex]
&+
\frac{2 t^2}{45 (t-1)^8 \,(t+1)^8} \,
\left[15 - 30\,t - 60\,t^2 + 150\,t^3 + 1547\,t^4 + 15956\,t^5 \right.
\nonumber\\[2ex]
& \quad\left. - 154368\,t^6 - 142420\,t^7 + 1166645\,t^8 + 357354\,t^9 
 - 2744516\,t^{10} - 276066\,t^{11} + 2046129\,t^{12}\right]\,.
\end{align}
The ``standard hypergeometric'' function which occurs in this 
expression, is encountered in various previous 
calculations~\cite{Pa1993,JePa1996} 
\begin{align}
\Phi(n, t) &=  {}_2F_1\left(1, - n t, 1 - n t, 
\left( \frac{1-t}{1+t} \right)^2 \right)
\nonumber\\[2ex]
&= -nt \sum_{k=0}^{\infty} 
\frac{\displaystyle \left( \frac{1-t}{1+t} \right)^{2 k}}{k-nt} \,.
\end{align}
The convergence of this series representation near $t=0$ 
is problematic, but it can be accelerated effectively using the 
combined nonlinear-condensation transformation (CNCT) 
described in Refs.~\cite{JeMoSoWe1999,AkSaJeBeSoMo2003}.
Using this method, we easily obtain the 
40-figure result
\begin{equation}
\label{res4P}
\ln k_0(n=4, l=1) =
-0.04195\,48945\,98085\,54867\,10375\,94335\,27134\,18570 \,,
\end{equation}
which is consistent with the 24-figure result given in 
Eq.~(64) of Ref.~\cite{AkSaJeBeSoMo2003} and with the 
27-figure result given in Table~III of Ref.~\cite{GoDr2000} for this state.

\begin{figure}
\includegraphics[width=0.7\linewidth]{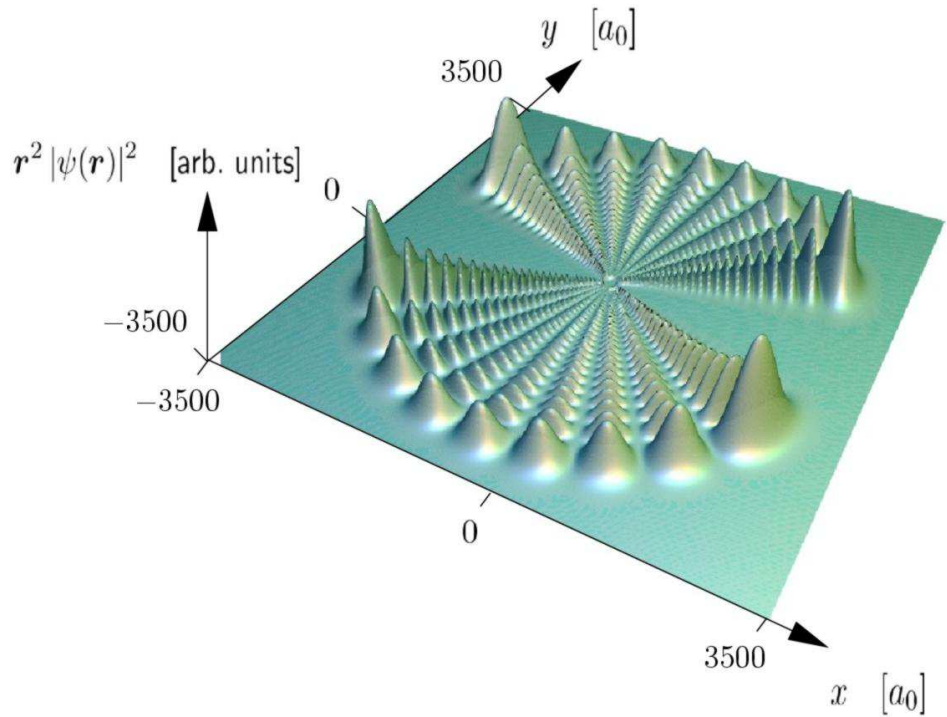}
\caption{\label{fig1} (color online.)
The number of terms in the radial wave function grows
with the principal quantum number $n$ and with the angular 
momentum defect $\zeta = n-l$. In order to illustrate 
this well-known fact, we here plot the radial
probability density of the 
state with quantum numbers $n=40$, $l=14$ and $m=6$,
in the plane of constant azimuth $\phi = 0$.
Here, $a_0$ denotes the Bohr radius
$a_0 = \hbar/(\alpha m c) = 0.529\,177\,2108(18) \times 
10^{-10}\,{\rm m}$~\cite{MoTa2005}. The Bethe 
logarithm for the state under discussion reads
$\ln k_0(n\!=\!40, l\!=\!14) = -0.418\,087\,713 \times 10^{-4}$.}
\end{figure}

The number of terms occurring in Eq.~(\ref{p4p}) is not excessive,
but it grows rapidly with the angular momentum defect 
$\zeta = n-l$. In addition, considerable numerical cancellation 
can occur from the typically alternating sign pattern of the polynomial 
that multiplies $\Phi(n, t)$.
For circular Rydberg states with $n = l+1$,
the analytic expressions obtained for $P(t)$ are most compact, 
and these states can be well treated using the 
integral representation,
up to very high principal quantum numbers.

We found that, using the integral representation,
numerically satisfactory results can be obtained for states with 
$\zeta = n-l < 5$, in the entire range $n \leq 200$. 
However, for $\zeta = n-l > 5$, the accuracy obtained 
using this method was not satisfactory, unless an excessively
accurate multiprecision arithmetic is used in intermediate steps
of the calculation. It is still possible to use the integral
representation for $n-l \leq 20$.  In this case, one does not have more
than 19 bound-state poles to subtract in forming the principal 
value in Eq.~(\ref{tlnk0}). 
As an alternative, one may deform the $t$-integration contour into 
the complex plane.
However, for $\zeta = n-l \leq 20$, we found the numerical
difficulties to be so severe that a different method 
of calculation appeared to be called for. 
The more complex structure of the wave function
with increasing $n-l$ is illustrated in Fig.~\ref{fig1}. 

However, before we resort to this different 
method in Sec.~\ref{srep} below,
we briefly dwell on the application of the integral 
representation to circular states with $n = l+1$.
It is possible to give
a general integral representation for the Bethe logarithm 
of a circular state with $l = n - 1$, $n > 1$.
This representation is a specialization 
of (\ref{tlnk0}) and reads
\begin{equation}
\label{bl}
\ln k_0(n,n-1) = -\frac34 \, \int_0^1 {\rm d} t\, f(t)\,.
\end{equation}
Here,
\begin{equation}
\label{ft}
f(t) = -\frac{1 - t^2}{n \, t^5} \,
\left[-\frac{2 t^2}{3 (1 - t^2)} + 
\sum^\infty_{k = 0} T(k,n,t)\right]\,.
\end{equation}
The term $T(k,n,t)$ is given by 
\begin{eqnarray}
\label{tknt}
T(k,n,t) &=& 
-\frac{2^{4 n} \, t^{2 n} \, (1 + t)^{-4 n}}
  {(2 n - 1) \, \Gamma(2 n)} \, 
\frac{1}{k!}\,
\left(\frac{1-t}{1+t}\right)^{2k}\,
\left[ \frac{(n - 1)\, t [{\cal P}(k,n,t)]^2\,
\Gamma(k + 2 n - 2)}{3 n \, (1 - t)^4 \, (1 - k - n + n \, t)} \right.
\nonumber\\[1ex]
& & \left. + \frac{4 t^3 \, \Gamma(k + 2 n + 2)}{3\,(1 + t)^4\, 
(-1 - k - n + n t)} \right]\,.
\end{eqnarray}
The polynomial ${\cal P}(k,n,t)$ reads
\begin{equation}
{\cal P}(k,n,t) = (k + n - 1)(2n - 1)(1 - t)^2 - 2 t k(k - 1)\,.
\end{equation}
The first term on the right-hand side of (\ref{tknt}) has 
a singularity at $t = (n-1)/n$, which corresponds to the 
decay into the lower-lying state with principal quantum 
number $n-1$ and orbital angular momentum quantum number $l=n-2$.
Selected numerical values for very highly excited hydrogenic 
states, obtained using the integral representation, are given in
Table~\ref{table1}.

\begin{center}
\begin{table}
\begin{center}
\begin{minipage}{16cm}
\caption{\label{table1}
Values of Rydberg state Bethe logarithms 
near $n\approx100$, for small angular momentum 
defect $\zeta = n-l$. The results displayed here
can be obtained using both the integral representation as well as 
the spectral representation of the Bethe logarithm,
and both methods were used in order to check the consistency of the results.}
\begin{tabular}{r@{\hspace{0.3cm}}r@{\hspace{0.3cm}}r@{\hspace{0.3cm}}r%
@{\hspace{0.3cm}}r@{\hspace{0.3cm}}r@{\hspace{0.3cm}}r}
\hline
\hline 
\rule[-3mm]{0mm}{8mm}
$\ln k_0(n,n-\zeta)$ &
\multicolumn{1}{c}{$\zeta=1$} &
\multicolumn{1}{c}{$\zeta=2$} &
\multicolumn{1}{c}{$\zeta=3$} &
\multicolumn{1}{c}{$\zeta=4$} \\
\hline
$n = 100$ & $  -0.583~308~014 \times 10^{-7}$ & $  -0.613~877~681 \times 10^{-7}$ & $  -0.645~944~796 \times 10^{-7}$ & $  -0.679~594~629 \times 10^{-7}$ \\ 
$n = 101$ & $  -0.566~008~997 \times 10^{-7}$ & $  -0.595~371~896 \times 10^{-7}$ & $  -0.626~158~390 \times 10^{-7}$ & $  -0.658~448~703 \times 10^{-7}$ \\ 
$n = 102$ & $  -0.549~387~309 \times 10^{-7}$ & $  -0.577~602~405 \times 10^{-7}$ & $  -0.607~171~570 \times 10^{-7}$ & $  -0.638~170~329 \times 10^{-7}$ \\ 
$n = 103$ & $  -0.533~410~121 \times 10^{-7}$ & $  -0.560~532~956 \times 10^{-7}$ & $  -0.588~944~368 \times 10^{-7}$ & $  -0.618~715~502 \times 10^{-7}$ \\ 
$n = 104$ & $  -0.518~046~496 \times 10^{-7}$ & $  -0.544~129~416 \times 10^{-7}$ & $  -0.571~439~188 \times 10^{-7}$ & $  -0.600~042~866 \times 10^{-7}$ \\ 
$n = 105$ & $  -0.503~267~262 \times 10^{-7}$ & $  -0.528~359~632 \times 10^{-7}$ & $  -0.554~620~643 \times 10^{-7}$ & $  -0.582~113~531 \times 10^{-7}$ \\ 
$n = 106$ & $  -0.489~044~896 \times 10^{-7}$ & $  -0.513~193~296 \times 10^{-7}$ & $  -0.538~455~408 \times 10^{-7}$ & $  -0.564~890~904 \times 10^{-7}$ \\ 
$n = 107$ & $  -0.475~353~416 \times 10^{-7}$ & $  -0.498~601~821 \times 10^{-7}$ & $  -0.522~912~079 \times 10^{-7}$ & $  -0.548~340~528 \times 10^{-7}$ \\ 
$n = 108$ & $  -0.462~168~279 \times 10^{-7}$ & $  -0.484~558~230 \times 10^{-7}$ & $  -0.507~961~045 \times 10^{-7}$ & $  -0.532~429~945 \times 10^{-7}$ \\ 
$n = 109$ & $  -0.449~466~295 \times 10^{-7}$ & $  -0.471~037~051 \times 10^{-7}$ & $  -0.493~574~371 \times 10^{-7}$ & $  -0.517~128~556 \times 10^{-7}$ \\ 
$n = 110$ & $  -0.437~225~533 \times 10^{-7}$ & $  -0.458~014~220 \times 10^{-7}$ & $  -0.479~725~687 \times 10^{-7}$ & $  -0.502~407~501 \times 10^{-7}$ \\ 
\hline
\hline
\end{tabular}
\end{minipage}
\end{center}
\end{table}
\end{center}

\begin{center}
\begin{table}
\begin{center}
\begin{minipage}{16cm}
\caption{\label{table2}
Sample values of Rydberg state Bethe logarithms 
near $n\approx200$, $0 \leq l \leq 3$.
All decimal figures shown are significant.
The values are consistent with a constant limit for $\ln k_0(n,l)$
as $n \to \infty$ for constant $l$.
In contrast to Table~\ref{table1}, the values displayed here
have been obtained exclusively using the spectral representation.
Note that the entries are labeled as $\ln k_0(n,l)$ in
contrast to the notation $\ln k_0(n,n-\zeta)$ used 
in Table~\ref{table1}.}
\begin{tabular}{r@{\hspace{0.3cm}}r@{\hspace{0.3cm}}r@{\hspace{0.3cm}}r%
@{\hspace{0.3cm}}r@{\hspace{0.3cm}}r@{\hspace{0.3cm}}r}
\hline
\hline 
\rule[-3mm]{0mm}{8mm}
$\ln k_0(n,l)$ & 
\multicolumn{1}{c}{$l=0$} & 
\multicolumn{1}{c}{$l=1$} & 
\multicolumn{1}{c}{$l=2$} & 
\multicolumn{1}{c}{$l=3$} \\
\hline
$n = 190$ & $   0.272~266~958 \times 10^{1}$ & $  -0.490~489~444 \times 10^{-1}$ & $  -0.993~712~588 \times 10^{-2}$ & $  -0.355~864~236 \times 10^{-2}$ \\ 
$n = 191$ & $   0.272~266~942 \times 10^{1}$ & $  -0.490~490~025 \times 10^{-1}$ & $  -0.993~716~023 \times 10^{-2}$ & $  -0.355~866~654 \times 10^{-2}$ \\ 
$n = 192$ & $   0.272~266~927 \times 10^{1}$ & $  -0.490~490~596 \times 10^{-1}$ & $  -0.993~719~406 \times 10^{-2}$ & $  -0.355~869~035 \times 10^{-2}$ \\ 
$n = 193$ & $   0.272~266~911 \times 10^{1}$ & $  -0.490~491~159 \times 10^{-1}$ & $  -0.993~722~737 \times 10^{-2}$ & $  -0.355~871~380 \times 10^{-2}$ \\ 
$n = 194$ & $   0.272~266~896 \times 10^{1}$ & $  -0.490~491~713 \times 10^{-1}$ & $  -0.993~726~017 \times 10^{-2}$ & $  -0.355~873~690 \times 10^{-2}$ \\ 
$n = 195$ & $   0.272~266~881 \times 10^{1}$ & $  -0.490~492~258 \times 10^{-1}$ & $  -0.993~729~247 \times 10^{-2}$ & $  -0.355~875~964 \times 10^{-2}$ \\ 
$n = 196$ & $   0.272~266~867 \times 10^{1}$ & $  -0.490~492~796 \times 10^{-1}$ & $  -0.993~732~428 \times 10^{-2}$ & $  -0.355~878~205 \times 10^{-2}$ \\ 
$n = 197$ & $   0.272~266~852 \times 10^{1}$ & $  -0.490~493~325 \times 10^{-1}$ & $  -0.993~735~562 \times 10^{-2}$ & $  -0.355~880~412 \times 10^{-2}$ \\ 
$n = 198$ & $   0.272~266~838 \times 10^{1}$ & $  -0.490~493~846 \times 10^{-1}$ & $  -0.993~738~649 \times 10^{-2}$ & $  -0.355~882~586 \times 10^{-2}$ \\ 
$n = 199$ & $   0.272~266~824 \times 10^{1}$ & $  -0.490~494~360 \times 10^{-1}$ & $  -0.993~741~690 \times 10^{-2}$ & $  -0.355~884~728 \times 10^{-2}$ \\ 
$n = 200$ & $   0.272~266~810 \times 10^{1}$ & $  -0.490~494~865 \times 10^{-1}$ & $  -0.993~744~687 \times 10^{-2}$ & $  -0.355~886~838 \times 10^{-2}$ \\ 
\hline
\hline
\end{tabular}
\end{minipage}
\end{center}
\end{table}
\end{center}

\begin{center}
\begin{table}
\begin{center}
\begin{minipage}{16cm}
\caption{\label{table3}
Values of Rydberg state Bethe logarithms 
near $n\approx200$, $100 \leq l \leq 103$.
This table complements Tab.~\ref{table1}, 
by investigating a range of quantum numbers where 
both the principal quantum number $n$ as well as the orbital 
angular momentum quantum number $l$ are large.
The angular momentum defect $\xi = n-l$ is also large 
for all states in this table.
As for the entries in Table~\ref{table2}, 
the values are consistent with a constant limit as $n \to \infty$
for given $l$.}
\begin{tabular}{r@{\hspace{0.3cm}}r@{\hspace{0.3cm}}r@{\hspace{0.3cm}}r%
@{\hspace{0.3cm}}r@{\hspace{0.3cm}}r@{\hspace{0.3cm}}r}
\hline
\hline 
\rule[-3mm]{0mm}{8mm}
$\ln k_0(n,l)$ &
\multicolumn{1}{c}{$l=100$} &
\multicolumn{1}{c}{$l=101$} &
\multicolumn{1}{c}{$l=102$} &
\multicolumn{1}{c}{$l=103$} \\
\hline
$n = 190$ & $  -0.108~830~510 \times 10^{-6}$ & $  -0.105~112~540 \times 10^{-6}$ & $  -0.101~547~415 \times 10^{-6}$ & $  -0.981~275~887 \times 10^{-7}$ \\ 
$n = 191$ & $  -0.109~118~840 \times 10^{-6}$ & $  -0.105~395~841 \times 10^{-6}$ & $  -0.101~825~818 \times 10^{-6}$ & $  -0.984~012~208 \times 10^{-7}$ \\ 
$n = 192$ & $  -0.109~403~831 \times 10^{-6}$ & $  -0.105~675~864 \times 10^{-6}$ & $  -0.102~101~004 \times 10^{-6}$ & $  -0.986~716~920 \times 10^{-7}$ \\ 
$n = 193$ & $  -0.109~685~537 \times 10^{-6}$ & $  -0.105~952~663 \times 10^{-6}$ & $  -0.102~373~023 \times 10^{-6}$ & $  -0.989~390~540 \times 10^{-7}$ \\ 
$n = 194$ & $  -0.109~964~013 \times 10^{-6}$ & $  -0.106~226~290 \times 10^{-6}$ & $  -0.102~641~927 \times 10^{-6}$ & $  -0.992~033~569 \times 10^{-7}$ \\ 
$n = 195$ & $  -0.110~239~309 \times 10^{-6}$ & $  -0.106~496~796 \times 10^{-6}$ & $  -0.102~907~766 \times 10^{-6}$ & $  -0.994~646~499 \times 10^{-7}$ \\ 
$n = 196$ & $  -0.110~511~476 \times 10^{-6}$ & $  -0.106~764~231 \times 10^{-6}$ & $  -0.103~170~590 \times 10^{-6}$ & $  -0.997~229~814 \times 10^{-7}$ \\ 
$n = 197$ & $  -0.110~780~565 \times 10^{-6}$ & $  -0.107~028~643 \times 10^{-6}$ & $  -0.103~430~446 \times 10^{-6}$ & $  -0.999~783~983 \times 10^{-7}$ \\ 
$n = 198$ & $  -0.111~046~625 \times 10^{-6}$ & $  -0.107~290~080 \times 10^{-6}$ & $  -0.103~687~382 \times 10^{-6}$ & $  -0.100~230~947 \times 10^{-6}$ \\ 
$n = 199$ & $  -0.111~309~701 \times 10^{-6}$ & $  -0.107~548~591 \times 10^{-6}$ & $  -0.103~941~443 \times 10^{-6}$ & $  -0.100~480~673 \times 10^{-6}$ \\ 
$n = 200$ & $  -0.111~569~843 \times 10^{-6}$ & $  -0.107~804~219 \times 10^{-6}$ & $  -0.104~192~674 \times 10^{-6}$ & $  -0.100~727~619 \times 10^{-6}$ \\ 
\hline
\hline
\end{tabular}
\end{minipage}
\end{center}
\end{table}
\end{center}

\begin{center}
\begin{table}
\begin{center}
\begin{minipage}{16cm}
\caption{\label{table4}
Values of Rydberg state Bethe logarithms 
near $n\approx200$, for the highest principal 
quantum numbers and angular momenta 
under investigation in this article. 
The angular momentum defect is small for the states listed in this Table.
Consequently, the states listed here
can be calculated using both the integral as well as the 
spectral representation.}
\begin{tabular}{r@{\hspace{0.3cm}}r@{\hspace{0.3cm}}r@{\hspace{0.3cm}}r%
@{\hspace{0.3cm}}r@{\hspace{0.3cm}}r@{\hspace{0.3cm}}r}
\hline
\hline 
\rule[-3mm]{0mm}{8mm}
$\ln k_0(n,l)$ &
\multicolumn{1}{c}{$l=196$} &
\multicolumn{1}{c}{$l=197$} &
\multicolumn{1}{c}{$l=198$} &
\multicolumn{1}{c}{$l=199$} \\
\hline
$n = 197$ & $  -0.753~369~175 \times 10^{-8}$ & - & - & - \\ 
$n = 198$ & $  -0.761~387~888 \times 10^{-8}$ & $  -0.741~963~223 \times 10^{-8}$ & - & - \\ 
$n = 199$ & $  -0.769~316~490 \times 10^{-8}$ & $  -0.749~821~211 \times 10^{-8}$ & $  -0.730~786~360 \times 10^{-8}$ & - \\ 
$n = 200$ & $  -0.777~156~469 \times 10^{-8}$ & $  -0.757~591~335 \times 10^{-8}$ & $  -0.738~487~630 \times 10^{-8}$ & $  -0.719~832~864 \times 10^{-8}$ \\ 
\hline
\hline
\end{tabular}
\end{minipage}
\end{center}
\end{table}
\end{center}

%
%
\section{Spectral Representation}
\label{srep}

The integral representation discussed in Sec.~\ref{trep}
reflects very closely the physics involved in the original
problem, by expressing the Bethe logarithm as an integral 
over the energy of a virtual photon.
For computational purposes, a different method can be more
effective, which relies on available analytic results
for transition matrix elements of 
discrete-discrete transitions~\cite{Go1929aop}, as cited in
Eq.~(63.2) of Ref.~\cite{BeSa1957},
and for discrete-continuum transitions [see Eq.~(6) of Ref.~\cite{ShBe1990}
or alternatively Ref.~\cite{KaLa1961}].
Note that in Eq.~(6) of Ref.~\cite{ShBe1990}, the argument 
of the arccot function in the exponential should be
replaced according to $n' n/n \to n'/n$, and 
that, as pointed out in Ref.~\cite{ShBe1991}, the 
continuum wave functions used in Ref.~\cite{ShBe1990} are
normalized to the energy scale, not to the momentum scale.
The latter fact implies that 
Eq.~(3b) of Ref.~\cite{ShBe1990} receives 
a correction according to Eq.~(3b) of Ref.~\cite{ShBe1991}.
For the transition matrix elements of selected 
low-lying states into the continuum, one may alternatively
use the formulas given in Ref.~\cite{St1930},
but one should be aware of multiplicative correction factors as pointed out
below Eq.~(11) of Ref.~\cite{BeBrSt1950}.

The spectral representation is based on the 
following formula for the Bethe logarithm,
\begin{align}
\label{bethelogp}
\ln k_0(n, l) = \frac{n^3}{2 (Z\alpha)^4\,m} \,
\sum^3_{i=1}
\left< nlm \left| \frac{p^i}{m} \,
\left( H_{\rm S} - E_n \right) \, 
\ln \left[ \frac{2 \left| H_{\rm S} - E_n \right|}{(Z\alpha)^2\,m} \right] \,
\frac{p^i}{m} \right| nlm \right>\,. 
\end{align}
Here, $H_{\rm S}$ is the Schr\"{o}dinger Hamiltonian
\begin{equation}
H_{\rm S} = \frac{\bm{p}^2}{2 m} - \frac{Z\alpha}{r}\,.
\end{equation}
This spectrum of this 
operator, as is well known, 
has a discrete part (bound states, $E_n < 0$), 
and a continuous part (continuum states, $E > 0$).
Note that the modulus $\left| H_{\rm S} - E_n \right|$ is
involved in Eq.~(\ref{bethelogp}). Otherwise,
the argument of the logarithm could become negative.
The specification of the modulus corresponds to the 
principal value prescription in Eq.~(\ref{tlnk0}).
Using the commutator relation $p^i = {\rm i} \, m\,
[H_{\rm S}, r^i]$, one may easily transform
Eq.~(\ref{bethelogp}) into
\begin{align}
\label{bethelogx}
\ln k_0(n, l) = \frac{n^3}{2 (Z\alpha)^4\,m} \,
\left< \phi \left| r^i \,
\left( H_{\rm S} - E_n \right)^3 \, 
\ln \left[ \frac{2 \left| H_{\rm S} - E_n \right|}{(Z\alpha)^2\,m} \right] \,
r^i \right| \phi \right>\,. 
\end{align}
We assume the wave functions of the continuous spectrum 
to be normalized according to 
\begin{equation}
\langle E l m | E' l' m'  \rangle = \delta(E - E') \,
\delta_{ll'}\, \delta_{mm'}\,.
\end{equation}
The discrete spectrum is normalized according to 
$\langle nlm | n'l'm' \rangle = \delta_{nn'}\, \delta_{ll'}\, \delta_{mm'}$.
One may then write down a spectral decomposition
of (\ref{bethelogp}),
\begin{align}
\ln k_0(n, l) \; =& \; \frac{n^3}{2 (Z\alpha)^4\,m} \,
\sum^3_{i=1}
\left< n l m\left| r^i \,
\left( H_{\rm S} - E_n \right)^3 \, 
\ln \left[ \frac{2 \left| H_{\rm S} - E_n \right|}{(Z\alpha)^2\,m} \right] \,
r^i \right| n l m \right>
\nonumber\\[2ex]
\; =& \; \frac{n^3}{2 (Z\alpha)^4\,m} \,
\sum_{n' = 0}^\infty\,\,
\sum_{l' = l\pm 1}\,\,
\sum_{m' = -l'}^{l'}\,\,
\sum^3_{i=1}
\left( E_{n'} - E_n \right)^3 \, 
\ln \left[ \frac{2 \left| E_{n'} - E_n \right|}{(Z\alpha)^2\,m} \right] \,
\left| \left< nlm \left| r^i \right| n'l'm' \right> \right|^2
\nonumber\\[2ex]
 & \; + \frac{n^3}{2 (Z\alpha)^4\,m} \,
\int_0^\infty {\rm d}E'\,\,
\sum_{l' = l\pm 1}\,\,
\sum_{m' = -l'}^{l'}\,\,
\sum^3_{i=1}
\left( E' - E_n \right)^3 \, 
\ln \left[ \frac{2 \left| E' - E_n \right|}{(Z\alpha)^2\,m} \right] \,
\left| \left< nlm \left| r^i \right| E' l'm' \right> \right|^2
\nonumber\\[2ex]
\; =& \; B + C\,.
\end{align}
Here, $B$ is the bound-spectrum contribution,
and $C$ stems from the continuum. Using the dipole 
selection rules, one immediately sees that the angular sums over 
$l'$ collapse to only two nonvanishing terms. We assume
the wave functions to have the structure
$\langle \bm{r} | nlm \rangle = 
R_{nl}(r) \, Y_{lm}(\theta,\phi)$
and $\langle \bm{r} | Elm \rangle = 
{\cal R}_{El}(r) \, Y_{lm}(\theta,\phi)$,
where $\theta$ and $\phi$ are the polar angles.
In terms of radial integrals,
the quantities $B$ and $C$ read
\begin{subequations}
\begin{align}
\label{brep}
B \; =& \; \frac{n^3}{2 (Z\alpha)^4\,m} \,
\left\{ \frac{l}{2 l + 1}\,
\sum_{n' = 0}^\infty\,\,
\left( E_{n'} - E_n \right)^3 \, 
\ln \left[ \frac{2 \left| E_{n'} - E_n \right|}{(Z\alpha)^2\,m} \right] \,
\left| \int_0^\infty {\rm d}r \, r^3 \, R_{nl}(r) \, R_{n'(l-1)}(r) \right|^2
\right.
\nonumber\\[2ex]
\; & 
\left. 
+ \frac{l + 1}{2 l + 1}\,
\sum_{n' = 0}^\infty\,\,
\left( E_{n'} - E_n \right)^3 \, 
\ln \left[ \frac{2 \left| E_{n'} - E_n \right|}{(Z\alpha)^2\,m} \right] \,
\left| \int_0^\infty {\rm d}r \, r^3 \, R_{nl}(r) \, R_{n'(l+1)}(r) \right|^2
\right\}
\end{align}
and
\begin{align}
\label{crep}
C \; =& \; \frac{n^3}{2 (Z\alpha)^4\,m} \,
\left\{ \frac{l}{2 l + 1}\, \int_0^\infty {\rm d}E'\,\,
\left( E' - E_n \right)^3 \, 
\ln \left[ \frac{2 \left| E' - E_n \right|}{(Z\alpha)^2\,m} \right] \,
\left| \int_0^\infty {\rm d}r \, r^3 \, R_{nl}(r) \, 
{\cal R}_{E'(l-1)}(r)  \right|^2
\right.
\nonumber\\[2ex]
 & \; \left. + \frac{l + 1}{2 l + 1}\,
\int_0^\infty {\rm d}E'\,\,
\left( E' - E_n \right)^3 \, 
\ln \left[ \frac{2 \left| E' - E_n \right|}{(Z\alpha)^2\,m} \right] \,
\left| \int_0^\infty {\rm d}r \, r^3 \, R_{nl}(r) \, 
{\cal R}_{E'(l+1)}(r)  \right|^2
\right\}\,.
\end{align}
\end{subequations}
The representations (\ref{brep}) and (\ref{crep})
involve only radial integrals; therefore
Eq.~(63.2) of Ref.~\cite{BeSa1957} and
Eq.~(6) of Ref.~\cite{ShBe1990} can be directly applied
(with the above mentioned correction in the 
argument of the arccot function).
In the calculation of $B$, the CNCT~\cite{JeMoSoWe1999,AkSaJeBeSoMo2003} 
was used in order to 
accelerate the convergence of the sum over $n'$.
For the evaluation of $C$, a simple Gaussian integration
was found to be appropriate after a suitable variable
change that maps the interval $E' \in (0,\infty)$ onto 
the compact interval $(0, 1)$.

Selected numerical values for very highly excited 
Rydberg states with principal quantum numbers $n \leq 200$, 
obtained using the spectral 
representation, can be  found in Tables~\ref{table2},~\ref{table3}, %
and~\ref{table4}. The values listed in Table~\ref{table2}
are consistent with the asymptotic expansions in 
Appendix~\ref{asymp}, as given in 
Eqs.~(\ref{sasymp}) and~(\ref{pasymp}),
and in Table~\ref{table5} (see also Ref.~\cite{Po1981}).
For circular states, the values given in  Table~\ref{table4}
confirm the asymptotic expansion in Eq.~(\ref{circasymp}).
Final numerical calculations were done using the high-performance
computing facilities of the Max Planck Institute for Nuclear
Physics in Heidelberg, and using a cluster
of IBM Thinkpad mobile workstations~\cite{Disclaimer}. 
Advantage has been taken of multiprecision 
libraries~\cite{Wo1988,Ba1990tech,Ba1993,Ba1994tech}.

%
%
\section{Conclusions}
\label{conclu}

We have presented the evaluation of Bethe logarithms
for all the 20100 hydrogenic states with principal quantum number
$n \leq 200$. Two methods have been used: the first 
method involves an integral 
representation which reflects the physics of the underlying 
phenomenon in a very direct manner, by expressing the 
Bethe logarithm in terms of an integral over the virtual
photon energy (see Sec.~\ref{trep}). 
The second method relies on a spectral decomposition of the 
Bethe logarithm (see Sec.~\ref{srep}). In that latter representation, the 
two distinct contributions from virtual discrete states
and virtual bound states can be clearly distinguished.
A third method, which has been used in exploratory work,
is briefly described in Appendix~\ref{drep}.
Selected numerical data are presented in 
Eq.~(\ref{res4P}) and in the 
Tables~\ref{table1}---\ref{table4}.
The full set of numerical values is available
at~\cite{HomeNISTHD}. 

Incidentally, we observe that for all states which simultaneously
fulfill $150 \leq n \leq 200$ and $l \geq 150$, the 
virtual bound states give by far the dominant contribution
to the Bethe logarithm; indeed, we have $|B/C| > 10^{10}$ for these
states 
[for the definition of $B$ and $C$ see Eqs.~(\ref{brep}) and~(\ref{crep})]. 
This observation is in sharp contrast to lower-lying states,
where $|B/C|$ is typically smaller than one. E.g., the 
ground state fulfills $|B/C| \approx 0.00483$.
We conclude that the contribution of virtual bound states as compared
to virtual continuum states is much more pronounced for 
Rydberg states as compared to lower-lying states,
an observation which might appear counter-intuitive 
at first glance.

\section*{Acknowledgments}

U.D.J.~acknowledges helpful conversations with Professor
Krzysztof Pachucki and support from the National Institute
of Standards and Technology during a number of research appointments.
Sabine Jentschura is acknowledged for carefully reading the manuscript,
and for extensive help in the numerical calculations regarding the 
comparison of the results of the current investigation to those
of Ref.~\cite{Po1981}.

\appendix

%
%
\section{Lattice Schr\"{o}dinger--Coulomb Propagator}
\label{drep}

The computational implementation
of the Schr\"{o}dinger propagator on a discrete
lattice has been discussed in Ref.~\cite{SaOe1989}.
It can lead to a computationally cheap 
evaluation procedure, provided that the 
required numerical accuracy is not excessive.
Numerical difficulties grow for higher excited states,
in both one- and two-loop quantum electrodynamic problems,
for of a number of reasons (more terms and in general
a more complex structure
of the wave function, more bound-state poles along the 
integration contours, more nodes of the wave function,
which translate into numerical cancellations, etc.).
While of course any computational method is expandable, 
we have found it difficult to control the accuracy 
of the Bethe logarithm, calculated using a 
discrete-lattice representation, for 
both for high $n$ as well as for high $\zeta = n-l$.
For high $n$, the wave function extends over many Bohr radii,
which necessitates an accurate representation of the 
Schr\"{o}dinger--Coulomb
propagator over an extended grid, which is difficult 
to achieve with a limited grid size.
For high $\zeta$, the difficulties are enhanced due to the 
oscillations of the wave function which 
necessitate an even more accurate representation
on the grid. 

In exploring this way of calculating the Bethe logarithm,
we found it instructive, however, to use routines 
which lead to an explicit diagonalization of the 
Schr\"{o}dinger--Coulomb
propagator on the grid.
The accuracy of the lowest virtual-state energy
eigenvalues is actually satisfactory,
while for higher excited virtual states, the eigenvalues
depart rapidly from the exact Schr\"{o}dinger
solution $E_n = -(Z\alpha)^2 m/(2 n^2)$.
This phenomenon has also been observed in the context
of basis-set calculations of relativistic effects in 
atoms which rely on $B$-spline techniques, see e.g. Ref.~\cite{JoBlSa1988}.
On a discrete lattice, one can in principle 
only obtain a discrete spectrum,
which for the Schr\"{o}dinger--Coulomb
propagator extends into the positive-energy domain.
It is then possible to use the 
eigenvalues and eigenvectors directly 
in order to evaluate the Bethe logarithm [in the sense 
of the bound-state contribution in Eq.~(\ref{brep})]. 
This statement remains true
although the higher excited virtual states ``energies''
on the lattice depart very much from the true
eigenvalues $E_n = -(Z\alpha)^2 m/(2 n^2)$ obtained for 
a Hamiltonian acting on $L^2(\mathbbm{R}^3)$.

Using the lattice representation, it is easily possible to obtain 
about 9 decimal figures for the $1S$ Bethe logarithm
on a lattice which extends to 20 Bohr radii and which 
is comprised of only 200 nodes. However, we have found it 
difficult to substantially enhance the accuracy,
for low-lying states, beyond 
20 figures, even if quadruple precision is used in the 
linear algebra libraries. This level of accuracy is of course dwarfed by 
other available methods, for the concrete problem at hand 
[see Eq.~(\ref{res4P})],
and therefore the lattice representation has not been pursued 
any further in the current context of Bethe logarithms for 
Rydberg states. However, we re-emphasize here that the lattice representation
can lead to a computationally very efficient evaluation
of matrix elements of the hydrogenic propagator,
a property which has become useful in the 
calculation of other quantum electrodynamic effects for lower-lying 
states with $n \leq 6$~\cite{PaJe2003,Je2004b60}.

\begin{center}
\begin{table}[htb]
\begin{center}
\begin{minipage}{16cm} 
\caption{\label{table5}
Numerical values of the limits $\ln k_0(\infty, l) 
\equiv \lim_{n \to \infty} \ln k_0(n, l)$
for $l=0,\dots,10$. The evaluation proceeds 
according to methods outlined in Ref.~\cite{Po1981};
the values communicated here are in agreement with and more 
accurate than those obtained for the range $l=0,\dots,7$
in Ref.~\cite{Po1981}.}
\begin{tabular}{r@{\hspace{0.3cm}}r}
\hline
\hline
\rule[-3mm]{0mm}{8mm}
l & \multicolumn{1}{c}{$\ln k_0(\infty,l)$} \\
\hline
0 &    2.722~654~335\\
1 &   -0.049~054~544\\
2 &   -0.009~940~457\\
3 &   -0.003~560~999\\
4 &   -0.001~663~771\\
5 &   -0.000~908~042\\
6 &   -0.000~548~999\\
7 &   -0.000~356~923\\
8 &   -0.000~244~981\\
9 &   -0.000~175~372\\
10 &  -0.000~129~830 \\
\hline
\hline
\end{tabular}
\end{minipage}
\end{center}
\end{table}
\end{center}

\begin{figure}
\includegraphics[width=0.5\linewidth]{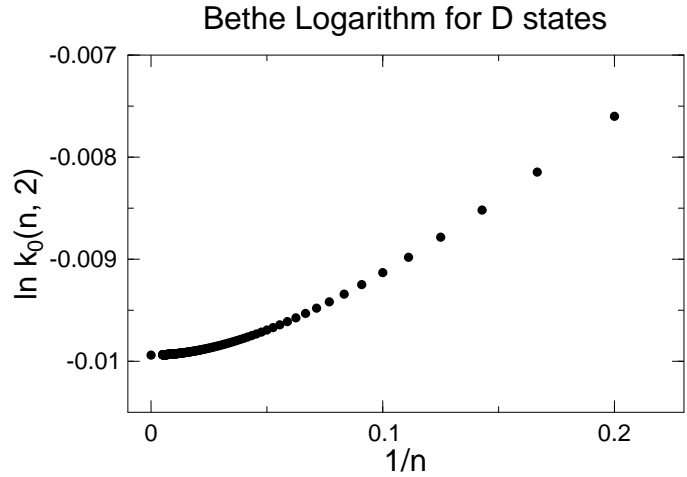}
\caption{\label{fig2} 
The Bethe logarithms for $l=2$ are in 
excellent agreement with their
asymptotic limit as $n\to \infty$. 
We plot here the values $\ln k_0(n, 2)$ as a function of $n^{-1}$.
For $n = 190,\dots, 200$, the values of 
$\ln k_0(n, 2)$ are given in Table~\ref{table2}
(see Ref.~\cite{HomeNISTHD} for
a complete list of all relevant values in the 
range $n\leq 200$). The limiting value at $n^{-1} = 0$,
which is $\ln k_0(\infty, 2) = -0.994~045~690 \times 10^{-2}$
(see also Table~\ref{table5}),
has been evaluated independently according to Ref.~\cite{Po1981}.}
\end{figure}

%
%
\section{Asymptotics Derived Previously}
\label{asymp}

Based on an extrapolation 
of the numerical data of Tab.~I of Ref.~\cite{JeEtAl2005}
to higher principal quantum numbers,
the following asymptotics
for the Bethe logarithm of circular Rydberg states had been 
obtained:
\begin{eqnarray}
\label{circasymp}
l^3 \times  \ln k_0(l+1, l) 
&\simeq&
-0.05685281(3) + \frac{0.0248208(6)}{l} 
+ \frac{0.03814(2)}{l^2}
- \frac{0.1145(5)}{l^3} 
+ \frac{0.166(3)}{l^4} 
- \frac{0.22(2)}{l^5} \,.
\end{eqnarray}
Here, terms of order~$l^{-k}$ with $k \geq 6$ are neglected.
For $S$ states, the following asymptotics had been obtained 
on the basis of the numerical data listed in Ref.~\cite{DrSw1990}:
\begin{eqnarray}
\label{sasymp}
\ln k_0(n, l=0) 
&\simeq&
2.72265434(5) + \frac{0.000000(5)}{n} 
+ \frac{0.55360(5)}{n^2}
- \frac{0.5993(5)}{n^3} 
+ \frac{0.613(7)}{n^4} 
- \frac{0.60(5)}{n^5} \,.
\end{eqnarray}
The corresponding expression for $P$ states reads
\begin{eqnarray}
\label{pasymp}
\ln k_0(n, l=1) 
&\simeq&
-0.0490545(1) + \frac{0.000000(5)}{n} 
+ \frac{0.20530(15)}{n^2}
- \frac{0.599(5)}{n^3} 
+ \frac{1.45(10)}{n^4} 
- \frac{3(1)}{n^5} \,.
\end{eqnarray}

In an apparently not very widely known paper~\cite{Po1981},
a method has been indicated for the evaluation of the Bethe logarithm 
in the limit of infinite principal quantum number $n= \infty$.
This method relies on an asymptotic expansion of the 
radial integrals that enter into Eqs.~(\ref{brep}) and (\ref{crep}),
in the limit of an infinite principal quantum number of the 
reference state. The resulting integrals are very slowly convergent,
but they permit a completely independent evaluation of 
$\ln k_0(\infty, l)$ which does not rely on an extrapolation
of data available for lower principal quantum numbers
[cf.~Eqs.~(\ref{circasymp}), (\ref{sasymp}) and (\ref{pasymp})]. 
Thus, the limiting values $\ln k_0(\infty, l)$ provide 
a sensitive independent cross-check of the numerical methods employed in 
the current calculation. 

In Ref.~\cite{Po1981}, the limits $\ln k_0(n=\infty, l)$ have been 
evaluated for $l=0,\dots,7$. Here, 
we generalize the treatment to the range $l=0,\dots,10$,
thereby confirming the asymptotic limits obtained in
Ref.~\cite{Po1981} (see Table~\ref{table5}).
A comparison to the numerical values obtained in the current 
investigation indicates excellent agreement with the 
asymptotic values as $n \to \infty$ (see Fig.~\ref{fig2}
for the case $l=2$). A full investigation of the asymptotic structure
of Bethe logarithms for large $n$, including a 
derivation of the subleading terms in the
expansion in powers of $n^{-1}$, would be very
interesting in its own right. Such an investigation 
would be facilitated by the availability of accurate 
numerical data over wide ranges of $n$ and $l$,
as obtained in the current investigation.


\begin{thebibliography}{10}
\bibitem{Disclaimer}
Certain commercial equipment, instruments, or materials are identified in this
  paper to foster understanding. Such identification does not imply
  recommendation or endorsement by the National Institute of Standards and
  Technology, nor does it imply that the materials or equipment identified are
  necessarily the best available for the purpose.
\end{thebibliography}
\end{document}